\documentclass{svjour2}
\smartqed
\usepackage{graphicx}
\usepackage{amssymb,latexsym,amsfonts,amsmath}

\journalname{Journal of Low Temperature Physics}
\begin{document}

\title{High frequency sound in superfluid $^3$He-B}

\author{J.P. Davis \and H. Choi \and J. Pollanen \and W.P. Halperin}

\institute{J.P. Davis \at
                 Department of Physics and Astronomy, Northwestern University,
Evanston, IL 60208, USA \\
                 Tel.: 780-248-1410\\
                 \email{jdavis@ualberta.ca}            \\
                \emph{Present address:} J.P. Davis,
                   	 Department of Physics, University of 
Alberta, Edmonton,
Alberta T6G 2G7, Canada
              \and
		W.P. Halperin \at
                 Department of Physics and Astronomy, Northwestern University,
Evanston, IL 60208, USA \\
                 Tel.: 847-491-3686\\
                 \email{w-halperin@northwestern.edu}
                       }
\date{Version \today}

\maketitle

\begin{abstract} We present measurements of the absolute phase velocity of
transverse and longitudinal sound in superfluid $^3$He-B at low temperature,
extending from the imaginary squashing mode to near pair-breaking.  Changes in
the transverse phase velocity near pair-breaking have been explained  in terms
of an order parameter collective mode that arises from $f$-wave pairing
interactions, the so-called $J=4^-$ mode.  Using these measurements, we
establish lower bounds on the energy gap in the B-phase.  Measurement of
attenuation of longitudinal sound at low temperature and energies far above the
pair-breaking  threshold, are in agreement with the lower bounds set on
pair-breaking.  Finally, we discuss  our estimations for the strength of the
$f$-wave pairing interactions and the Fermi  liquid parameter, $F_4^s$.

\end{abstract}

\section{Introduction}
\label{intro}

\begin{figure}[t]
\centerline{\includegraphics[width=4.0in]{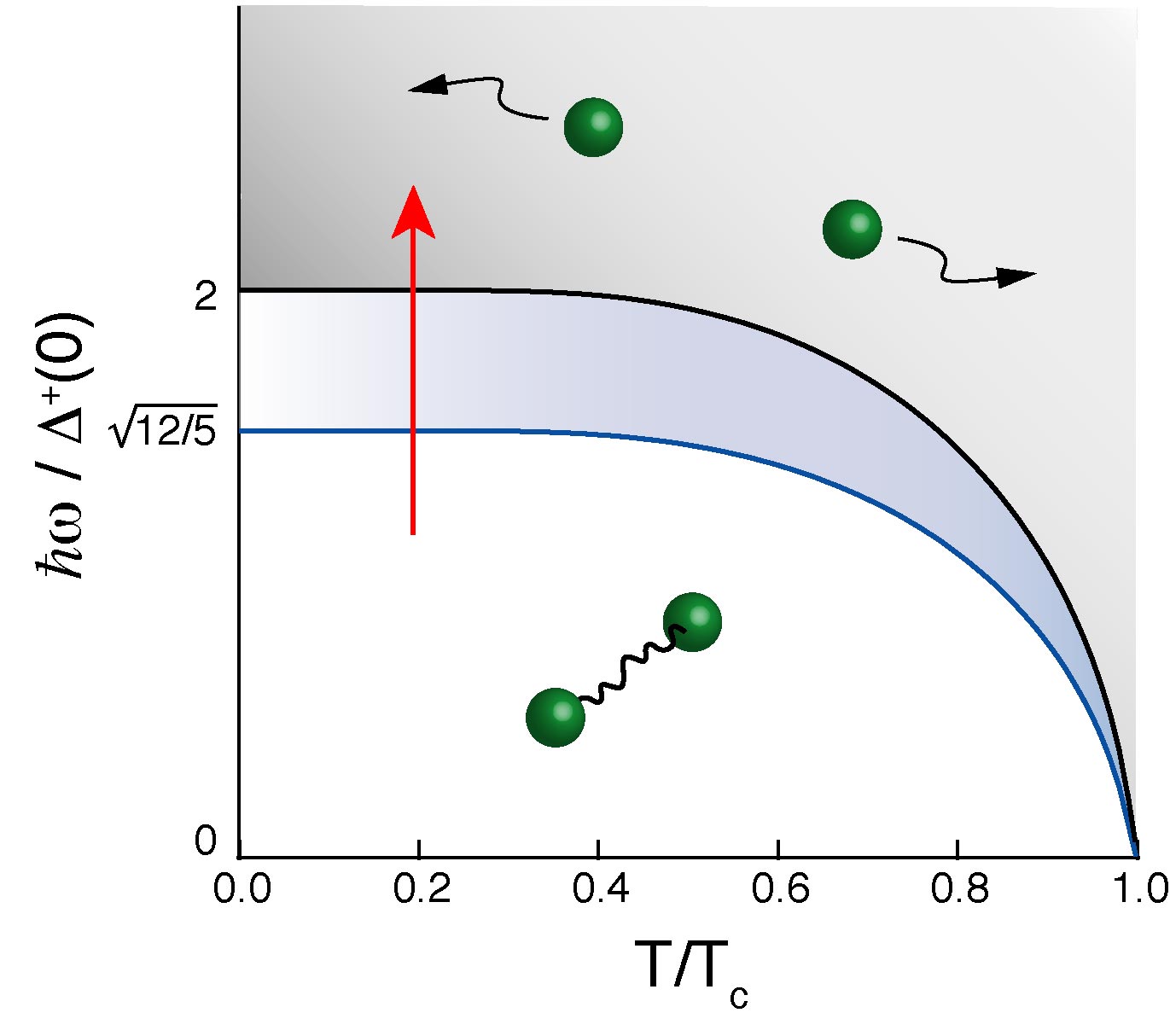}}
\caption{\label{fig1}(Color on-line).  Schematic of the threshold energy for
pair-breaking (black curve) and the ISQ-mode (blue curve) as a function of
temperature normalized to the transition temperature.  The shaded  blue region
is the only region in which transverse sound propagates.  Decreasing the
pressure at low temperature (pressure sweep technique) sweeps the acoustic
wave energy relative to the gap energy, represented by the red arrow.  The
shaded black region is the particle-hole continuum identified schematically by
independent quasiparticles. Below the black curve, they are shown as a bound
Cooper pair.}
\end{figure}

	In 1957 Landau \cite{Lan57} predicted the existence of new 
collective modes of
a degenerate Fermi liquid  in the high frequency and low temperature limits
where local thermal equilibrium cannot be established. One of these so-called
zero sound modes, the longitudinal mode, was observed by Abel {\it et al.}
\cite{Abe66} in 1966.   The second predicted zero sound mode was transverse
sound.  It is intriguing that the conditions on shear stiffness that are
required for propagation  of a transverse wave might be satisfied in a liquid,
and in particular in the Fermi liquid
$^3$He. Nonetheless, the second mode predicted by Landau has never been
observed \cite{Hal90}. Moores and Sauls \cite{Moo93} pointed out that
transverse sound might propagate at low temperatures in the B-phase of
superfluid
$^3$He as a consequence of its coupling to the imaginary squashing mode (ISQ),
a well-established collective mode of the order parameter.  They predicted a
significant increase in the velocity owing to an off-resonant coupling with
the ISQ which would significantly enhance the observability of transverse
sound.  The phenomenon was observed by Lee {\it et al.} \cite{Lee99}
exploiting an acoustic Faraday effect to unambiguously demonstrate that the
propagating mode they discovered had transverse polarization.  According to
the theory and subsequent experimental  work
\cite{Moo93,Lee99,Dav06,Dav08,Dav08b}, we know that this coupled transverse
mode propagates only at  frequencies between that of the ISQ-mode, shown as
the blue curve in Fig.~\ref{fig1}, and  pair breaking, the black curve. In
the present work we investigate the region near  the pair breaking threshold
in superfluid $^3$He exploiting both transverse and  longitudinal sound modes
using acoustic cavity inerference methods.

	We recently reported observation of a new  collective mode near the
pair-breaking energy,
$2\Delta$, based on anomalies in phase velocity, attenuation and acoustic
birefringence of transverse sound \cite{Dav08b}. The additional measurements
we present here are made  possible by improvements in acoustic techniques,
notably acoustic cavity interferometry and pressure sweeps at low
temperature.  Our measurements of this mode establish a lower bound for the
energy gap,
$\Delta$,  since the energy, $\hbar\omega$, of the 2$\Delta$-mode is close to,
but below, the pair-breaking threshold,
$2\Delta$.  Previous reports of the pair-breaking threshold based on
longitudinal sound attenuation \cite{Mov90,Mas00}  are difficult to reconcile
with theory since they give values (at low pressure) less than BCS
weak-coupling.  In the pressure range from 1 to 20 bar, we find that the gap
always lies above weak-coupling values.  Additionally, we compare measurements
of longitudinal sound with
    transverse sound.  We find that longitudinal sound continues to propagate at
energies far above $2\Delta$,  albeit with significantly high attenuation.
   From longitudinal sound attenuation we extract the threshold energy for
pair-breaking at 3.7 bar, consistent with the lower bounds established by the
$2\Delta$-mode.

	Selection rules for the coupling between transverse sound and 
order parameter
collective modes provide the basis for our identification of the
2$\Delta$-mode as having total angular momentum $J \geq 4$.   A $J=4$ mode has
been predicted  to exist if there is an attractive, sub-dominant
$f$-wave pairing interaction \cite{Sau86}.  There have been predictions for
sub-dominant pairing interactions in high $T_c$ superconductors
\cite{Fog97} and in superfluid $^3$He \cite{Sau86,Sau81}.  Superfluid $^3$He is
a spin triplet, $p$-wave pairing condensate where the possible role of
$f$-wave interactions has been the subject of considerable interest both
experimentally and theoretically.   Several experiments, including magnetic
susceptibility
\cite{Fis87,Fis88} and spectroscopy of the ISQ and real squashing (RSQ)
collective modes
\cite{Hal90,Dav06,Dav08,Sau82,Fra89,Fra90}, have been analyzed to try and
determine the $f$-wave pairing interaction, as well as Fermi liquid
interactions, in an effort to predict the $J=4$ mode frequency.  However, the
results of these different analyses are ambiguous owing to imprecision of the
Fermi liquid parameters, $F^{a,s}_{2}$, as well as non-trivial strong coupling
contributions to the ISQ-mode frequencies \cite{Dav08}.  The strength of the
$f$-wave pairing interactions can be estimated from the $2\Delta$-mode energies
in terms of the theory
\cite{Sau81} if assumptions are made about the high order Fermi liquid
parameter, $F_4^s$.  Conversely, we can estimate $F_4^s$ using independent
values for the $f$-wave pairing strength, taken from measurements of another
order parameter collective mode, the RSQ.  It is traditional to refer to these
modes by their total angular momentum, $J$, and to indicate explicitly their
parity under particle-hole conversion.  Then the ISQ is called the
$J=2^-$ mode, the RSQ is the $J=2^+$ mode and the new mode, as predicted,
would be
$J=4^-$. Particle-hole symmetry is relevant to the strength of the coupling of
order parameter collective modes to sound \cite{Hal90,McK90}. For example, for
perfect particle-hole symmetry, which is the case of even parity, there can be
no coupling to the RSQ-mode.

\section{Longitudinal and Transverse Acoustic Response}

	The experimental arrangement is described in 
Refs.~\cite{Dav08,Dav08b}.  We
form an acoustic cavity with a spacing of $D = 31.6 \pm 0.1~\mu$m between an
\emph{AC}-cut transducer (6 MHz fundamental frequency) and an  optically
polished quartz reflector.  The temperature is held fixed near $\approx
0.5$~mK and the pressure is swept while keeping the acoustic frequency,
$\omega$, constant.  The energy gap, $\Delta$, is dependent on the pressure and
therefore a decreasing  pressure sweep changes the relative energy of the
acoustic waves to the energy gap, as well as to the order parameter collective
modes, shown schematically by the red arrow in Fig.~\ref{fig1}.  Specifically,
we use the weak-coupling-plus energy gap of Rainer and Serene \cite{Rai76},
referenced to the Greywall temperature scale \cite{Gre86}.  The low temperature
pressure sweep method results in crossing the $2\Delta$-mode at
$T / T_c = 0.29$ at a pressure of 19.4 bar (170 MHz), up to $T / T_c  = 0.53$
for 1.2 bar (76 MHz).

\begin{figure}[t]
\centerline{\includegraphics[width=4.5in]{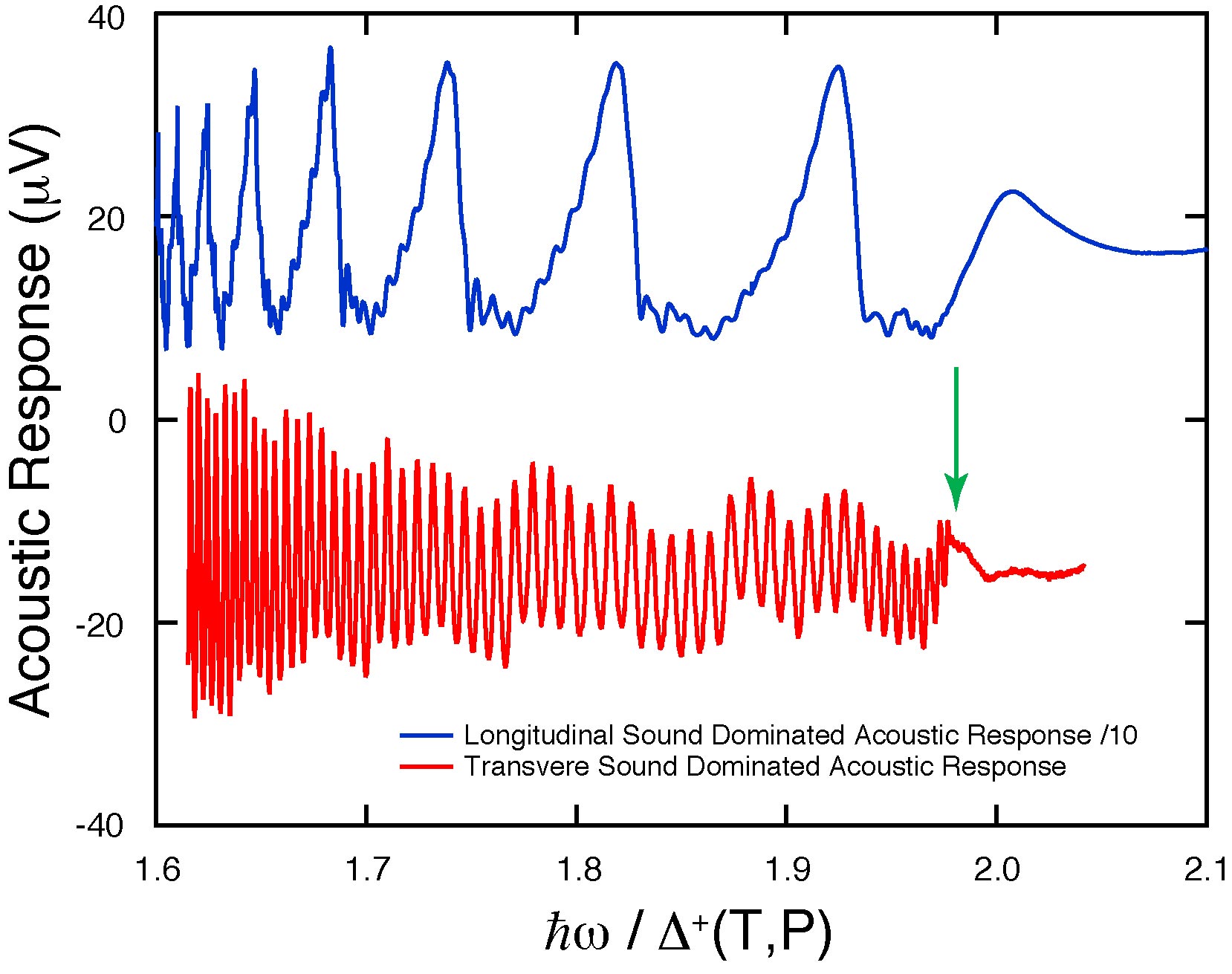}}
\caption{\label{fig2}(Color on-line).  Transverse acoustic response
oscillations (lower red trace) at $\approx 550~\mu$K and longitudinal acoustic
response oscillations (upper blue trace) at $\approx 600~\mu$K as a function
of energy normalized to the weak-coupling-plus gap.  The background has been
subtracted from the transverse trace to facilitate comparison.  The frequency
of  transverse sound in the lower trace is 88 MHz and is 99.8 MHz for the
longitudinal sound in the upper trace.  The green arrow indicates the position
of the $2\Delta$-mode, which is below $2\Delta^+(0)$ at the pressure
corresponding to this trace.  For the upper trace a modulation of approximately
10\% corresponds to transverse sound.  Conversely, longitudinal
sound is weakly generated in the lower trace.}
\end{figure}

	The acoustic response, $V_Z$, was detected using a cw-impedance bridge
\cite{Ham94}, which is sensitive to changes in the standing wave condition of
the sound wave in the cavity through  changes in the electrical  impedance of
the transducer.  The transverse acoustic response is given by
\begin{equation}\label{VZ}  V_Z=a+b\cos\theta\sin\bigg(\frac{2D\omega}{c} +
\phi \bigg).
\end{equation}	  The angle $\theta$ is the direction of polarization of the
sound wave at the surface of the  transducer relative to the  polarization of
the shear transducer; $c$ is the sound velocity and $\phi$ is a  fixed phase
that depends on experimental conditions.  This is superposed on a slowly
varying background, $a$, not shown in Fig.~\ref{fig2}.  Hence, the acoustic
response is sensitive to: the phase velocity, proportional to the period
between the acoustic response oscillations; attenuation, increasing with
decreasing amplitude of the acoustic response oscillations and proportional to
$b$; and the polarization of the transverse sound wave.   We work  from 76 to
170 MHz, the
$13^{th}$ to
$29^{th}$ harmonics of the transducer.  The $17^{th}$ harmonic of our
\emph{AC}-cut transducer, generates primarily longitudinal sound, likely from a
slightly miscut transducer.  This allows us to perform longitudinal sound
measurements with the high resolution in frequency and the sensitivity of a
shear transducer and directly compare longitudinal and transverse acoustic
response traces.  The acoustic response for longitudinal sound is similar to
that of Eq.~\ref{VZ}, but without the polarization ($\cos\theta$) factor and
with $c$ corresponding to the velocity of longitudinal sound.

We show in Fig.~\ref{fig2} representative energy (pressure) sweeps using both
transverse and longitudinal acoustic response as a function of energy
normalized to the weak-coupling-plus gap
\cite{Hal90,Rai76}.  The green arrow indicates the energy of the
$2\Delta$-mode,
$\hbar\Omega_{2\Delta}$, in the transverse sound trace.  The $2\Delta$-mode was
identified by changes in the transverse phase velocity, attenuation  and
acoustic birefringence \cite{Dav08b}.  Further analysis of the attenuation of
transverse sound is discussed elsewhere \cite{Dav08c}.  We do not see a
dramatic change in the phase velocity of the longitudinal trace near
pair-breaking in Fig.~\ref{fig2} that would indicate coupling to this, or any
other, collective mode.  The small oscillations in the longitudinal trace are
from transverse acoustic response that is generated in addition to
longitudinal sound.

	For each full oscillation in the acoustic response, as in the 
lower trace in
Fig.~\ref{fig2}, one half wavelength of the transverse standing wave  has
entered or left the cavity as described by Eq.~\ref{VZ}.  Therefore the
velocity difference between each maximum (minimum) corresponds to a half
wavelength and  we convert the period of the acoustic response oscillations
into the change in the  phase velocity as a function of
pressure, $\Delta c$, where $(\Delta c/c)^{-1} =
\frac{D\omega}{\pi c} -1$.  Absolute values for the transverse sound phase
velocity, $c_t = c$, can then be obtained by fixing the velocity at one
particular frequency, which we take to be the phase velocity approaching the
ISQ-mode, through comparison with calculations based on the
    theoretical dispersion relation for transverse sound \cite{Moo93}.  This
dispersion relation in the long wavelength limit and zero magnetic field
\cite{Moo93,Sau99} is given by:
\begin{equation}\label{dispersion}
        \frac{\omega^2}{q^2 v_{F}^2} = \Lambda_{0} +
\Lambda_{2^{-}}\frac{\omega^{2}}{\omega^{2}-\Omega_{2^{-}}^{2}-\frac{2}{5}q^{2}v_{F}^{2}},
\end{equation} where $v_{F}$ is the Fermi velocity and $q$ is the complex
wavevector,
\begin{equation}\label{q} q = k + i\alpha,
\end{equation}
$k$ is the real wavevector, $\alpha$ is the attenuation, and the phase velocity
is $c_t =
\omega/k$.  The ISQ-mode frequency closely follows the temperature and pressure
dependence of the energy gap, $\Delta(T,P)$:
\begin{equation}\label{freq}
\Omega_{2^{-}}(T,P) = a_{2^{-}}(T,P)\Delta(T,P),
\end{equation} where $a_{2^{-}} \approx \sqrt{12/5}$.  For the  precise values
of
$a_{2^{-}}$ we use those  determined experimentally in
Refs.~\cite{Dav06,Dav08}.  The first term on the right hand side of
Eq.~\ref{dispersion} is the quasiparticle background, the contribution to the
dispersion in the absence of off-resonant coupling to the ISQ-mode,
\begin{equation}\label{L0}
\Lambda_{0} =
\frac{F_1^{s}}{15}(1-\lambda)\Big(1+\frac{F_{2}^{s}}{5}\Big)/\Big(1+\lambda\frac{F_{2}^{s}}{5}\Big).
\end{equation} The second term on the right hand side of Eq.~\ref{dispersion}
gives the off-resonant coupling of transverse sound to the ISQ with strength:
\begin{equation}\label{L2}
\Lambda_{2^{-}} =
\frac{2F_1^{s}}{75}\lambda\Big(1+\frac{F_{2}^{s}}{5}\Big)^{2}/\Big(1+\lambda\frac{F_{2}^{s}}{5}\Big).
\end{equation} Here we have included all Fermi liquid terms, $F_l^s$, for
$l\leq2$.   The Tsuneto function, $\lambda(\omega,T)$, can be thought of as a
frequency dependent superfluid stiffness \cite{Moo93} and is numerically
calculated using the weak-coupling-plus gap
\cite{Rai76}, $\Delta^{+}(T,P)$, tabulated by Halperin and Varoquaux
\cite{Hal90}.

\begin{figure}[t]
\centerline{\includegraphics[width=4.5in]{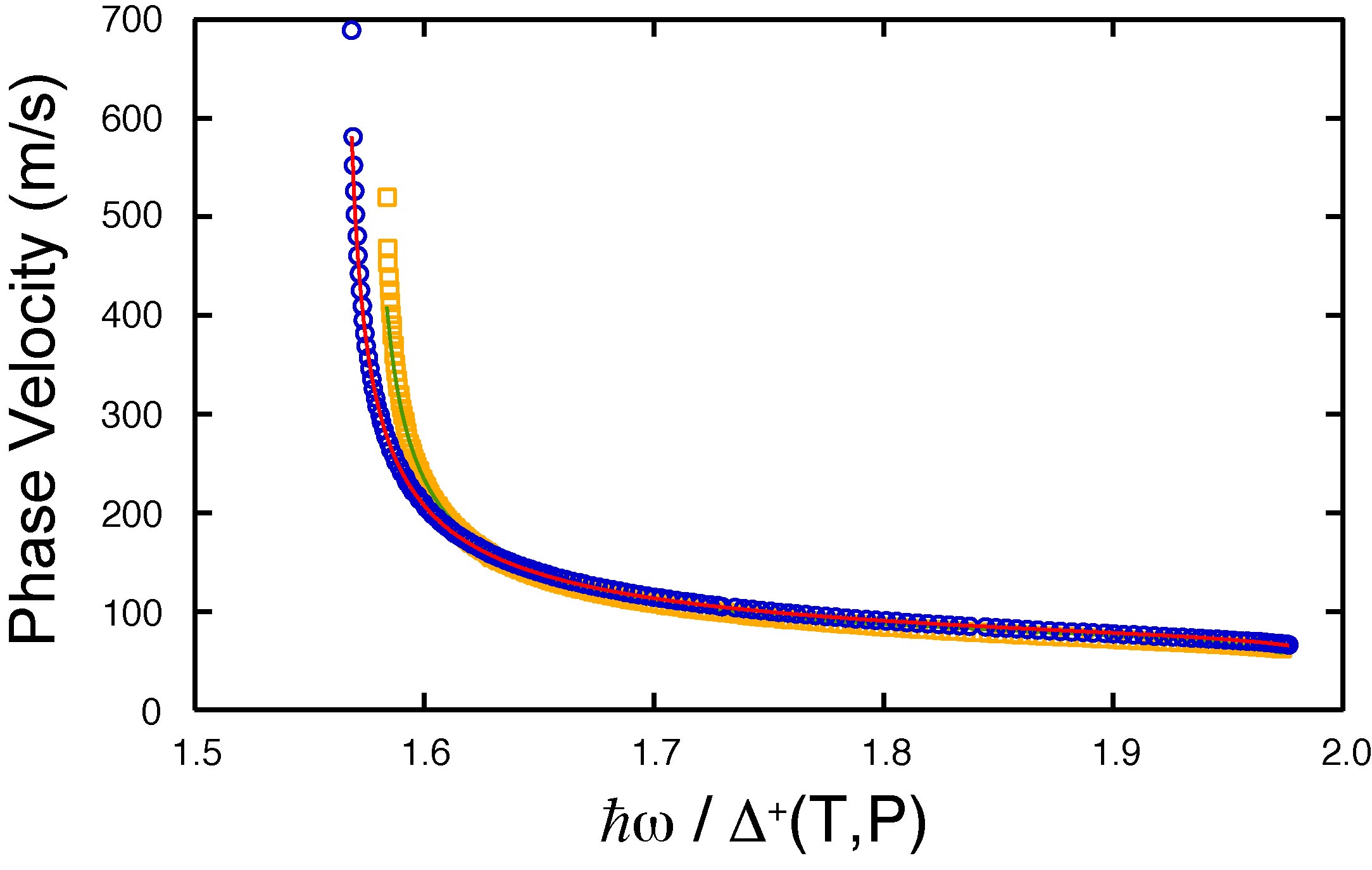}}
\caption{\label{Fig3}(Color on-line).  Phase velocity of transverse sound as a
function of energy normalized to the weak-coupling-plus gap at low temperature
($\approx 550~\mu$K).  The blue circles are at 88 MHz and the orange  squares
are at 111.5 MHz.  The red curve is the theoretical phase velocity for 88 MHz
with the new dispersion term and the green curve is for 111.5 MHz.}
\end{figure}

	The agreement between the above theoretical dispersion 
relation and the data
is excellent for energies below $\sim 1.8\Delta^+$.  To fit the sound velocity
above this energy we add a term to the dispersion, Eq.~\ref{dispersion}, which
corresponds to coupling of the
$2\Delta$-mode to transverse sound, with coupling strength,
$\Lambda_{2\Delta}$:
\begin{equation}\label{newmode}
    \frac{\omega^2}{q^2 v_{F}^2} = \Lambda_{0} +
\Lambda_{2^{-}}\frac{\omega^{2}}{\omega^{2}-\Omega_{2^{-}}^{2}-\frac{2}{5}q^{2}v_{F}^{2}}
+\Lambda_{2\Delta}\frac{1}{\omega^2 - \Omega_{2\Delta}^2}.
\end{equation}  This form neglects the width of the  collective modes, as well
as dispersion corrections to the $2\Delta$-mode energy, since it has not yet
been calculated.  The calculated velocity from this phenomenological dispersion
relation fits the data precisely, shown in Fig.~\ref{Fig3}, with coupling
strength, $\Lambda_{2\Delta}= 0.18$ for the 88 MHz data (red curve) and
$\Lambda_{2\Delta}= 0.15$ for the 111.5 MHz data (green curve).  At the higher
frequency (which corresponds to higher pressures) the transverse sound velocity
is slightly lower over most of the energy range.  This is in contrast with the
longitudinal sound velocity, $c_l$, which increases with increasing pressure
\cite{Hal90}.  Near pair-breaking $c_t=66.1$~m/s at 2.4 bar and 88 MHz, and
$c_t=62$~m/s at 5.6 bar and 111.5 MHz.  These values scale with the Fermi
velocities of 54.8 m/s and 50.3 m/s at 2.4 and 5.6 bar respectively, being
$\approx 22\%$ larger than the Fermi velocity near the frequency $\omega
\approx 2\Delta/\hbar$ in both cases.

\begin{figure}[t]
\centerline{\includegraphics[width=3.0in]{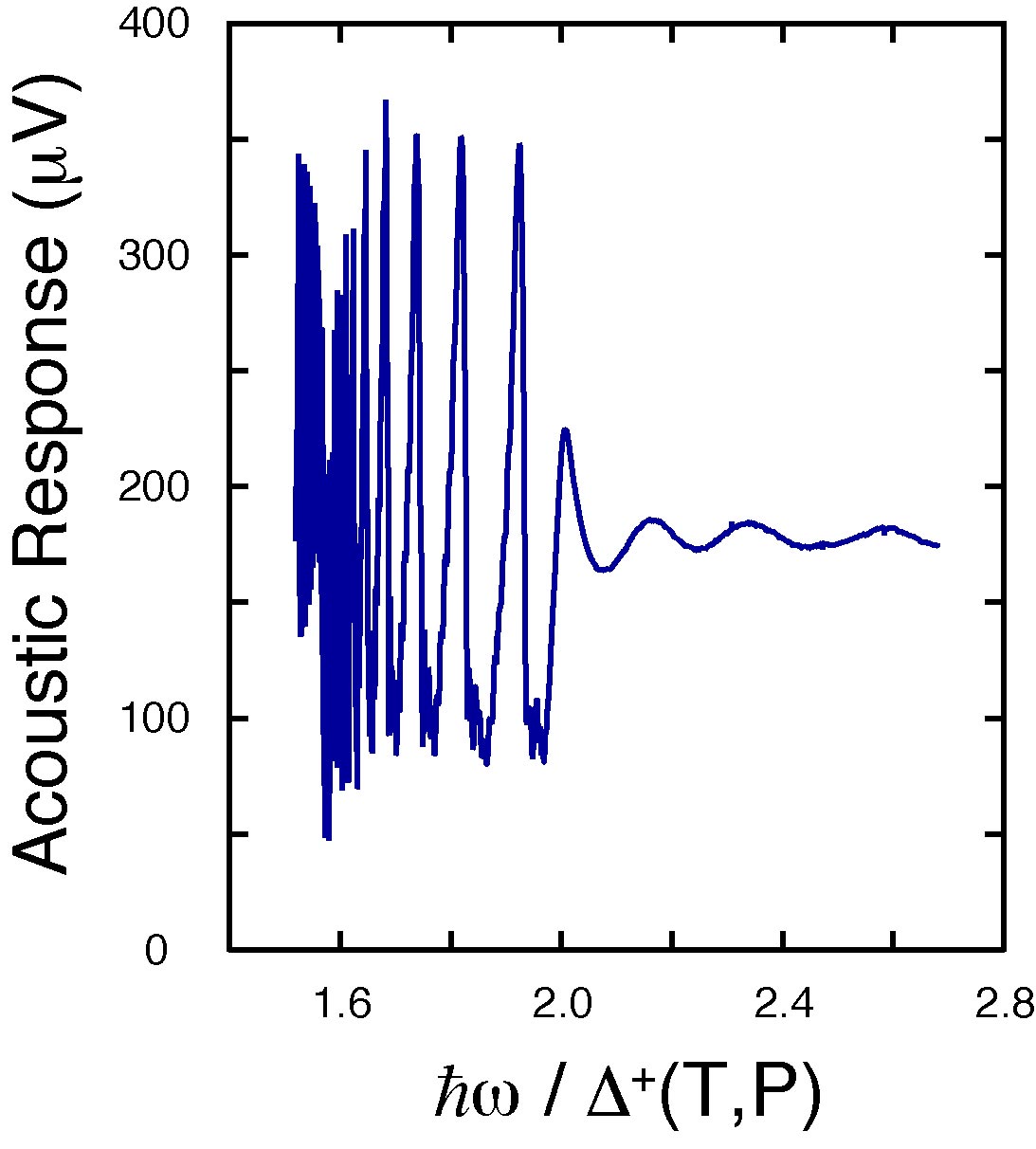}}
\caption{\label{fig4}(Color on-line).  Longitudinal acoustic response
oscillations as a function of energy normalized to the  weak-coupling-plus gap
at low temperature ($\approx 600~\mu$K) at 99.8 MHz.  The amplitude of  the
oscillations decreases above $\approx 2\Delta^+$, but they continue to the
highest energies, indicating that longitudinal sound propagates well into the
particle-hole continuum.}
\end{figure}

	In addition to coupling between transverse sound and the ISQ, 
there is also a
coupling between longitudinal sound and the ISQ.  Similar to the transverse
case, this gives an increased longitudinal sound velocity and increased
attenuation near the ISQ.  The dispersion relation for longitudinal sound in
superfluid $^3$He is given by \cite{Moo93}
\begin{equation}\label{LSdispersion}
        \frac{\omega^2}{q^2c_{1}^{2}} =1 +
\frac{4}{15}\Big(\frac{qv_F}{\omega}\Big)^{2}(1-\lambda) +
\frac{8}{75}\Big(\frac{qv_F}{\omega}\Big)^{2}\lambda\bigg[\frac{\omega^2}{\omega^2
- \Omega_{2^{-}}^{2} - \frac{7}{15} q^{2}v_{F}^{2}}\bigg],
\end{equation} where $c_1$ is the hydrodynamic sound velocity, $c_l =
\omega/k$ is the phase velocity and $q=k+i\alpha$.  Note that there are a few
significant differences between Eq.~\ref{LSdispersion} and
Eq.~\ref{dispersion}.  One very important difference is the first term on the
right side of Eq.~\ref{LSdispersion} that results in real valued solutions at
frequencies below the ISQ where longitudinal sound is allowed to propagate.
Additionally we find that there continues  to  be acoustic response
oscillations in the longitudinal trace over all energies that we could measure
even above pair-breaking, shown in Fig.~\ref{fig4}.
\begin{figure}[t]
\centerline{\includegraphics[width=4.0in]{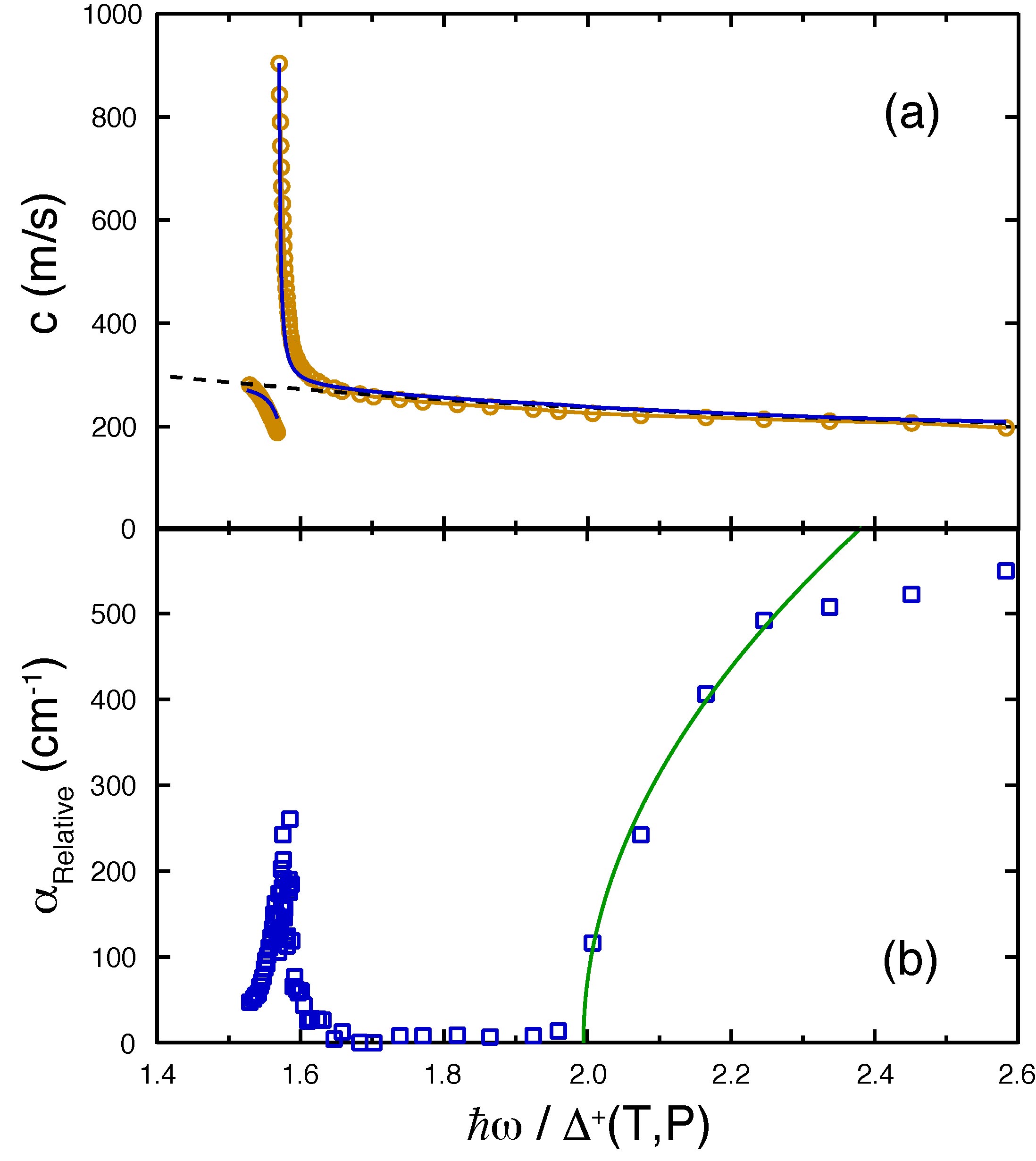}}
\caption{\label{LSbothCandAtten}(Color on-line).  Longitudinal sound velocity
(a) and attenuation (b) as a function of energy normalized to the
weak-coupling-plus gap at low temperature ($\approx 600~\mu$K) at 99.8 MHz.  In
the upper panel the theoretical sound velocity is given by the blue curve and
the data by the gold circles.  The pressure dependence of the  longitudinal
sound velocity in the absence of coupling to the ISQ-mode is given by the
dashed black curve.  In the lower panel the relative attenuation of
longitudinal sound is given as the blue squares.  The green curve is a fit to
Eq.~\ref{LSatten}, giving $2\Delta_{pair-breaking} = (1.994 \pm
0.006)\Delta^{+}$.}
\end{figure}

	The longitudinal acoustic response oscillations can be 
converted to  velocity
and attenuation as was done for transverse sound \cite{Dav08b,Dav08c}.  This is
shown in Fig.~\ref{LSbothCandAtten}.  The sound velocity,
Fig.~\ref{LSbothCandAtten}a, calculated from the theoretical dispersion
relation, Eq.~\ref{LSdispersion}, is given by the blue curve.  The data is then
fixed to the theoretical velocity near the ISQ-mode, thus determining the
longitudinal sound velocity at all other energies (gold circles).  The theory
and data are in good agreement over the entire energy range.  The pressure
dependence of the longitudinal sound velocity in the normal state, $c_1$ as
tabulated in Ref.~\cite{Hal90}, is given as the dashed black curve in
Fig.~\ref{LSbothCandAtten}a.  It is interesting to note that the dominant
contribution to the energy dependence of the longitudinal sound  velocity, at
all energies other than near the ISQ-mode, comes from its dependence  on
pressure during the pressure sweep.   Moreover, there seems to be no strong
effect on the longitudinal sound velocity near the pair-breaking energy where
we  have evidence for a new mode from transverse sound, possibly because of
the low resolution in energy in this measurement.  This is the first
measurement of the  phase velocity of longitudinal sound for such high
attenuation, which has been facilitated in our case by using a narrow cavity
and acoustic interference techniques.

	Furthermore, we can determine the relative attenuation of 
longitudinal sound
from the amplitude of  the acoustic response oscillations, shown in
Fig.~\ref{LSbothCandAtten}b.  It is reasonable to expect that  the attenuation
of longitudinal sound is nearly zero away from the ISQ-mode and pair-breaking
\cite{Lin87}, which would mean that the relative attenuation in
Fig.~\ref{LSbothCandAtten}b is nearly the  absolute attenuation.  The energy
resolution of this attenuation data is rather poor owing to the short acoustic
path length, but extends to
$\approx10 \times$ the attenuation of previous measurements, which allows us to
determine the attenuation at energies far above pair-breaking.  Unlike the
longitudinal sound velocity, the attenuation shows a dramatic effect near the
threshold energy.  It is expected \cite{McK90} that the
attenuation from pair breaking behaves as
\begin{equation}\label{LSatten}
\alpha \sim \sqrt{\hbar \omega - 2\Delta_{pair-breaking}}.
\end{equation} Fitting our data to this relation, the green curve in
Fig.~\ref{LSbothCandAtten}b, yields \\ $2\Delta_{pair-breaking} = (1.994 \pm
0.006)\Delta^{+}$.  At even higher energies, the attenuation deviates  from
this relation, which is not understood.  The nature of sound propagation deep
into the particle-hole continuum and at low temperatures is not
well-established, perhaps in part since there have been no experimental
results until this work. In this spirit we note that the experimental
conditions exceed the  quantum limit described by Landau \cite{Lan57} 
in his discussion of
zero sound propagation in the normal Fermi liquid.  In our case we have
$\hbar\omega / 2\pi k_B T \sim 1.3$. It is in this low temperature 
limit, so far
unexplored theoretically at high frequency in superfluid $^3$He, that the
quantum mechanical form of the quasiparticle distribution function should be
used. In the normal Fermi liquid it results in sound attenuation, even at zero
temperature.
\section{Lower Limits on the Pair-Breaking Energy}

	The single most fundamental quantity in the thermodynamic 
description of both
superconductors and superfluid $^3$He is the gap energy, $\Delta(T)$.
Understanding the gap energy of superfluid $^3$He is vital for correct
interpretation of collective mode frequencies, which scale with the
gap, as well as virtually all static properties of the superfluid.  The energy
gap is the energy per quasiparticle required to dissociate a Cooper pair into
its constituent quasiparticles.  Anisotropy in the gap as well as strong
coupling corrections to the free energy can in principle increase the gap
energy above $\Delta(0) = 1.764k_bT_c$.  Acoustic measurements of the energy
gap are based on the principle that its signature is a sharp onset of
attenuation associated with pair-breaking. Previous measurements of the pair
breaking energies using longitudinal sound have indicated anomalously low
values for the energy gap, below the 1.764$k_{B}T_{c}$ of BCS theory
\cite{Mov90,Mas00}.  In order to investigate this surprising result, we use the
$2\Delta$-mode as a lower bound on the pair-breaking energy of superfluid
$^3$He.  By doing this, we establish lower limits on the energy of the gap,
which we find to be above the weak-coupling BCS theory at all pressures.

	In this light, the results of Movshovich \emph{et al.} 
\cite{Mov90}  and
Masuhara
\emph{et al.}
\cite{Mas00} from longitudinal sound measurements, are particularly unsettling
since both of these experiments yield gap energies below $\Delta_{BCS}$, at low
pressures.  To date, these two experiments are the only ones to have measured
the gap at low temperatures, below $T / T_c < 0.5$.  Other experiments
\cite{Lin87,Gia80,Mei83} measured the pair breaking edge using attenuation of
longitudinal sound near
$T_c$.  Adenwalla \emph{et al.} \cite{Ade89} used longitudinal acoustics to
identify a feature with $2\Delta$ down to $T / T_c  \approx 0.6$.  These
experiments were consistent with the weak-coupling-plus model
\cite{Rai76} when combined with the temperature scale of Greywall \cite{Gre86}
but are not sensitive tests of the low temperature limit of the
weak-coupling-plus theory \cite{Rai76}.

\begin{figure}[t]
    \centerline{\includegraphics[width=4.5in]{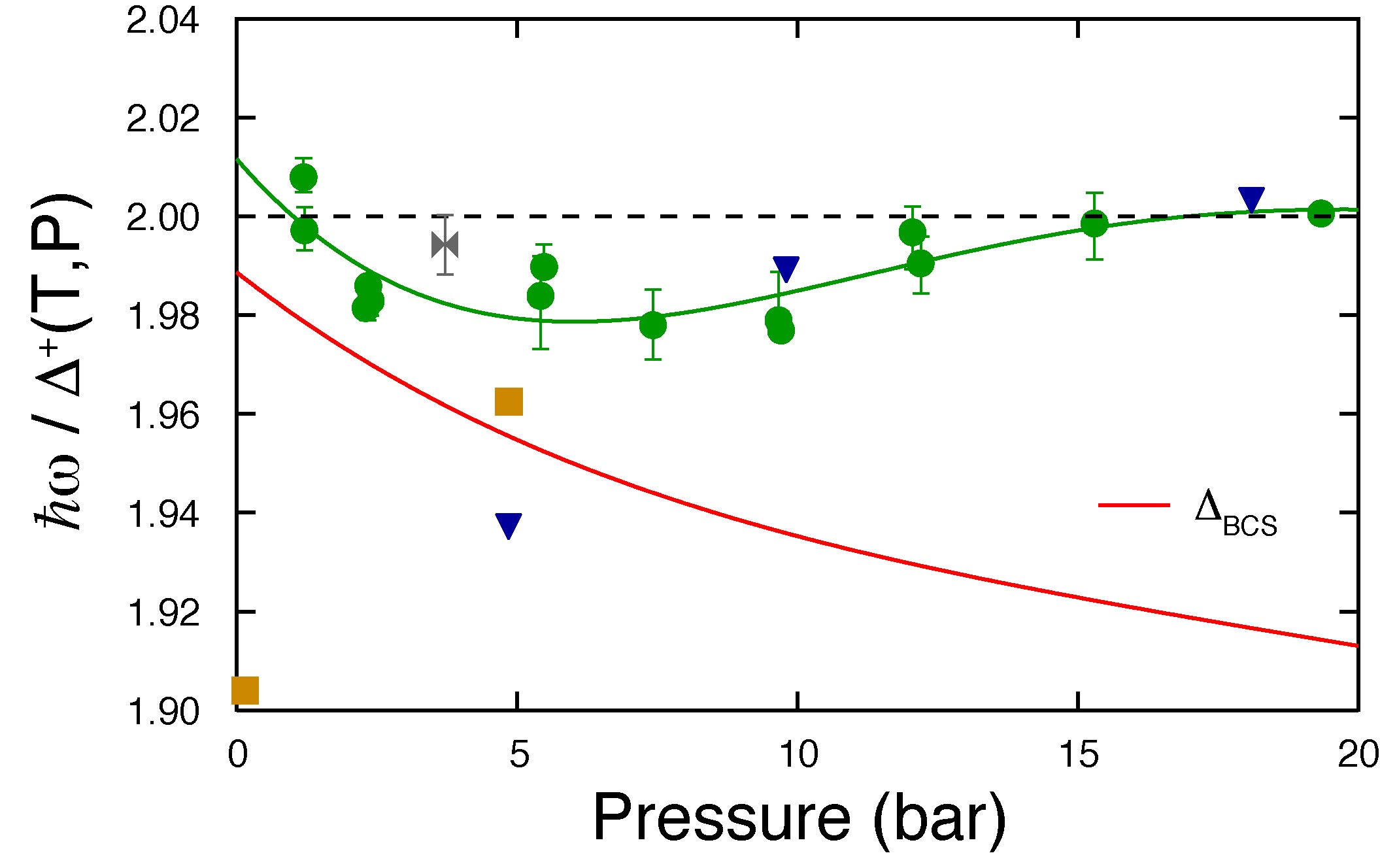}}
\caption{\label{fig6}(Color on-line).  Measurements of
$\hbar\Omega_{2\Delta}$, green circles, normalized to the gap at zero
temperature in the weak-coupling-plus model as a function of pressure.  The
green line is a fit to the data, Eq.~\ref{J4}.  The grey bow-tie is from the
attenuation of longitudinal sound.  The black dashed line is the
weak-coupling-plus gap
\cite{Rai76}.  The red line is the BCS weak-coupling gap
\cite{Hal90}.  The blue downward triangles are the data of Movshovich \emph{et
al.} and the gold squares are the data of Masuhara \emph{et al.}  We find an
upturn at low pressures for the minimum value for $2\Delta$, as opposed to the
precipitous drop to lower values for the data of Movshovich \emph{et al.} and
Masuhara \emph{et al.}}
\end{figure}

	Our measurement of $\hbar\Omega_{2\Delta}$, using transverse 
acoustic response
is the first experiment using transverse sound to shed light on the gap.  The
off-resonant coupling of transverse sound to the ISQ-mode is extinguished above
pair-breaking and transverse sound no longer propagates by this mechanism
\cite{Moo93}.  This is expected to behave like a step-function, sharply
increasing the attenuation and making transverse sound well-suited  for
measuring pair-breaking.  The only problem with this scenario is the
appearance of the
$2\Delta$-mode \cite{Dav08b}.  Nonetheless, using the $2\Delta$-mode as a
\emph{lower bound}, we can draw conclusions about the pair-breaking energy.
Comparing the transverse and  longitudinal traces, Fig.~\ref{fig2}, it is clear
that transverse sound does not propagate above pair-breaking in our acoustic
cavity.  It is widely believed that transverse sound should propagate in the
normal state
\cite{Lan57}, but we have found that the attenuation is too high for us to
observe standing waves with the acoustic path length used here,
$2D = 63.1~\mu$m.

	Fig.~\ref{fig6} shows the results of our 
$\hbar\Omega_{2\Delta}$ measurements,
normalized to the weak-coupling-plus gap, as a function of pressure.  We find
$\hbar\Omega_{2\Delta}$ is within $\approx 1\%$ of the expected value of the
weak-coupling-plus gap for $\lesssim 20$ bar and our interpretation is that the
gap energy \emph{never} falls below $\Delta_{BCS}$.  Our measurement of the
pair-breaking threshold energy extracted from longitudinal sound attenuation is
shown as the grey bow-tie in Fig.~\ref{fig6} and is consistent, in that it lies
at an energy above the measurements of
$\hbar\Omega_{2\Delta}$.  There is good agreement at pressures $\geq 10$ bar
between our measurements and those of Movshovich
\emph{et al.} \cite{Mov90}, but in our results there is a clear upturn at the
lowest pressures in stark contrast to the experiments of both Movshovich
\emph{et al.} \cite{Mov90} and Masuhara
\emph{et al.} \cite{Mas00}.  It is possible that Movshovich \emph{et al.} are
detecting an increased attenuation coming from the feature observed by Ling
\emph{et al.}  \cite{Lin87} that couples to longitudinal sound and resides near
$2\Delta$ in a magnetic field, although its exact origin is unknown
\cite{Mck93,Ash96}, since their experiment involved sweeping the magnetic field
to move the gap energy through their acoustic spectrum.  The results  of
Masuhara
\emph{et al.} are not subject to this source of ambiguity and it is difficult
for us to reconcile their experiment with ours.

\section{Extraction of $f$-Wave Pairing and Fermi Liquid
Interactions}\label{parameters}

	Predictions for the $J=4^-$ mode frequency, $\Omega_{4^-}$, 
depend on the
existence of an attractive $f$-wave pairing interaction \cite{Sau86},  as well
as the unknown Fermi liquid parameter, $F^{s}_{4}$.  Parameterization of the
$f$-wave pairing strength is given by
\begin{equation}\label{x3} x_3^{-1} = \frac{1}{v_1^{-1} -v_3^{-1}},
\end{equation} where $v_1$ and $v_3$ are the pairing potentials due to
$p$-wave and $f$-wave interactions respectively.  Attractive $f$-wave pairing
interactions correspond to negative values for
$x_3^{-1}$, and $x_3^{-1}$ is zero if the $f$-wave pairing interaction is zero.
Repulsive interactions give a positive $x_3^{-1}$.  The predicted
$J=4^-$ mode frequency in terms of $x_3^{-1}$ and
$F^{s}_{4}$ is given by \cite{Sau81}:
\begin{equation}\label{predJ4}
\Big(1+\frac{1}{9}F_4^s\lambda\Big) +
x_3^{-1}\lambda\bigg[\frac{\Omega_{4^-}^2}{4\Delta^2} -
\frac{5}{9} + \frac{5}{81}F_4^s\Big(\frac{\Omega_{4^-}^2}{4\Delta^2} -
1\Big)\lambda\bigg]=0,
\end{equation} for the case when $x_5^{-1}=0$ (zero $h$-wave pairing
interactions).

\begin{figure}[t]
       \centerline{\includegraphics[width=4.5in]{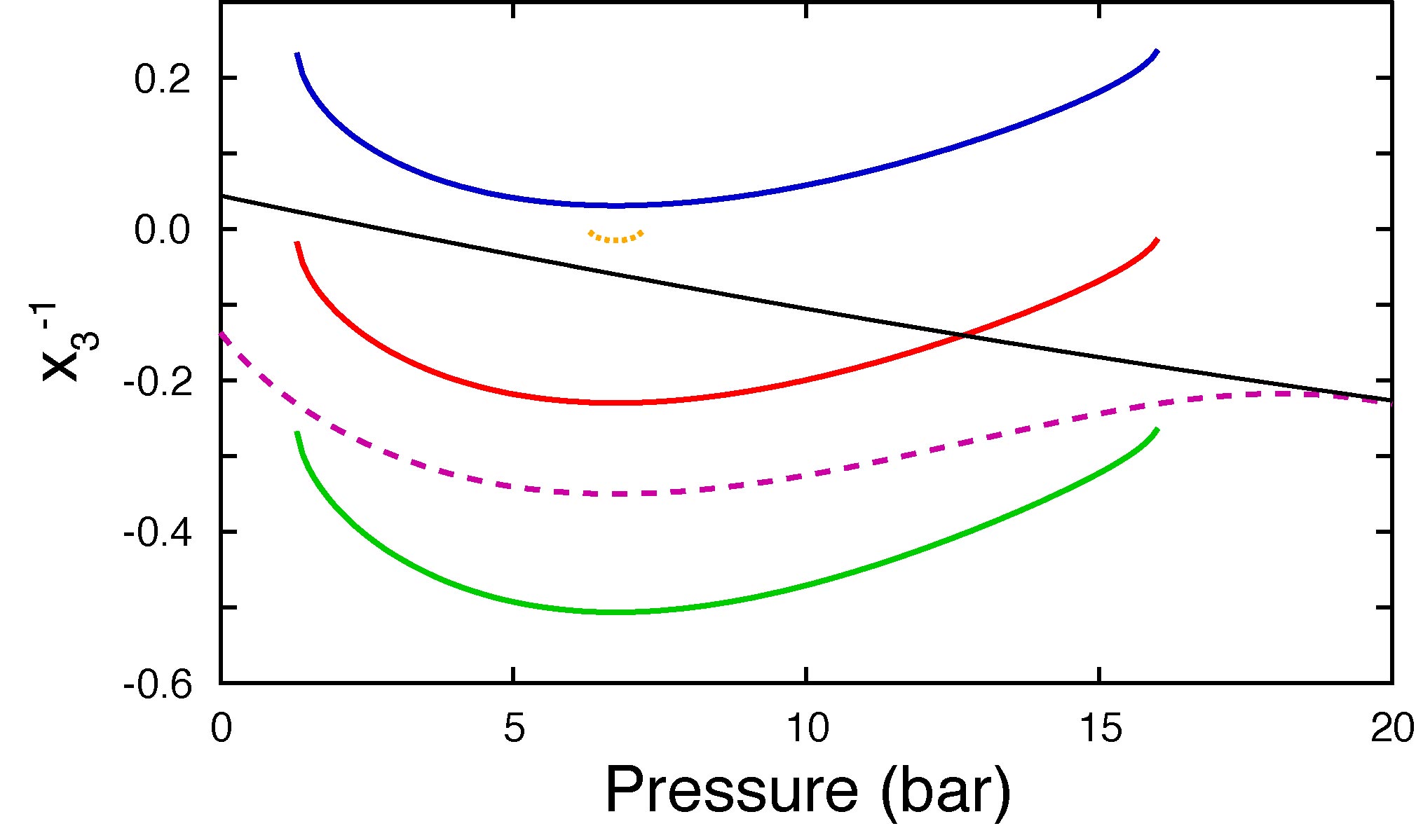}}
\caption{\label{x3inv}(Color on-line).  $f$-wave pairing interaction strength
from $2\Delta$-mode energies using Eq.~\ref{predJ4}.  The red curve uses
$F_4^s = 0$, the blue curve $F_4^s = -1$ and the green curve $F_4^s = +1$. The
purple dashed curve results from $F_4^s = 0$ and increasing $T_c$ (and
therefore $\Delta^+$) by 1\%, and the orange dotted curve results from
$F_4^s = 0$ and decreasing $T_c$ by 1\%.  Whenever $\Omega_{4^-} \geq
2\Delta^+$ the Tsuneto function is imaginary and there are no solutions to
Eq.~\ref{predJ4}.  The black line represents the the values of $x_3^{-1}$ from
Fraenkel \emph{et al.} \cite{Fra89}.}
\end{figure}

Evaluation of the strength of $f$-wave pairing interactions from the measured
values for the energy of the $2\Delta$-mode is complicated by the fact that
there is no experimental information on $F^{s}_{4}$.  Nonetheless, we can
evaluate Eq.~\ref{predJ4}, for various constant values of $F^{s}_{4}$.
Specifically  in the case where $F_4^s=0$ we can rewrite Eq.~\ref{predJ4} as
\begin{equation}\label{x3inverse} x_3^{-1} =
\bigg[\lambda\Big(\frac{\Omega_{4^-}^2}{4\Delta^2} -
\frac{5}{9}\Big)\bigg]^{-1}.
\end{equation} For $\Omega_{4^-}$ in Eq.~\ref{x3inverse}, we use our  values
from the
$2\Delta$-mode \cite{Dav08b} given up to 20 bar by:
\begin{multline} \label{J4}
\hbar\Omega_{2\Delta}(T,P) = 2\Delta^+(T,P) (1.006 - 6.7\times10^{-3}P +
8.7\times10^{-4}P^2
\\-3.94\times10^{-5}P^3 + 6\times10^{-7}P^4).
\end{multline} The values of $x_3^{-1}$ extracted from this analysis are
presented in Fig.~\ref{x3inv}.  When $\Omega_{4^-} \geq 2\Delta^+$ the Tsuneto
function is imaginary and there are no solutions to Eq.~\ref{predJ4}. The
values of $x_3^{-1}$ are sensitive to the energy of the collective mode
normalized to the pair-breaking energy, $\Omega_{4^-}/2\Delta^+$.  Since the
energy gap is proportional to $T_c$, uncertainty in the absolute temperature
will affect these values.  The uncertainty in $T_c$ was estimated by  Greywall
to be $\pm1\%$
\cite{Gre86} and we show the result of adjusting $T_c$ by $\pm1\%$ as  the
purple dashed and orange dotted curves respectively.  The effect of $T_c$ on
$\lambda$ was taken into account.

\begin{figure}[t]
       \centerline{\includegraphics[width=4.5in]{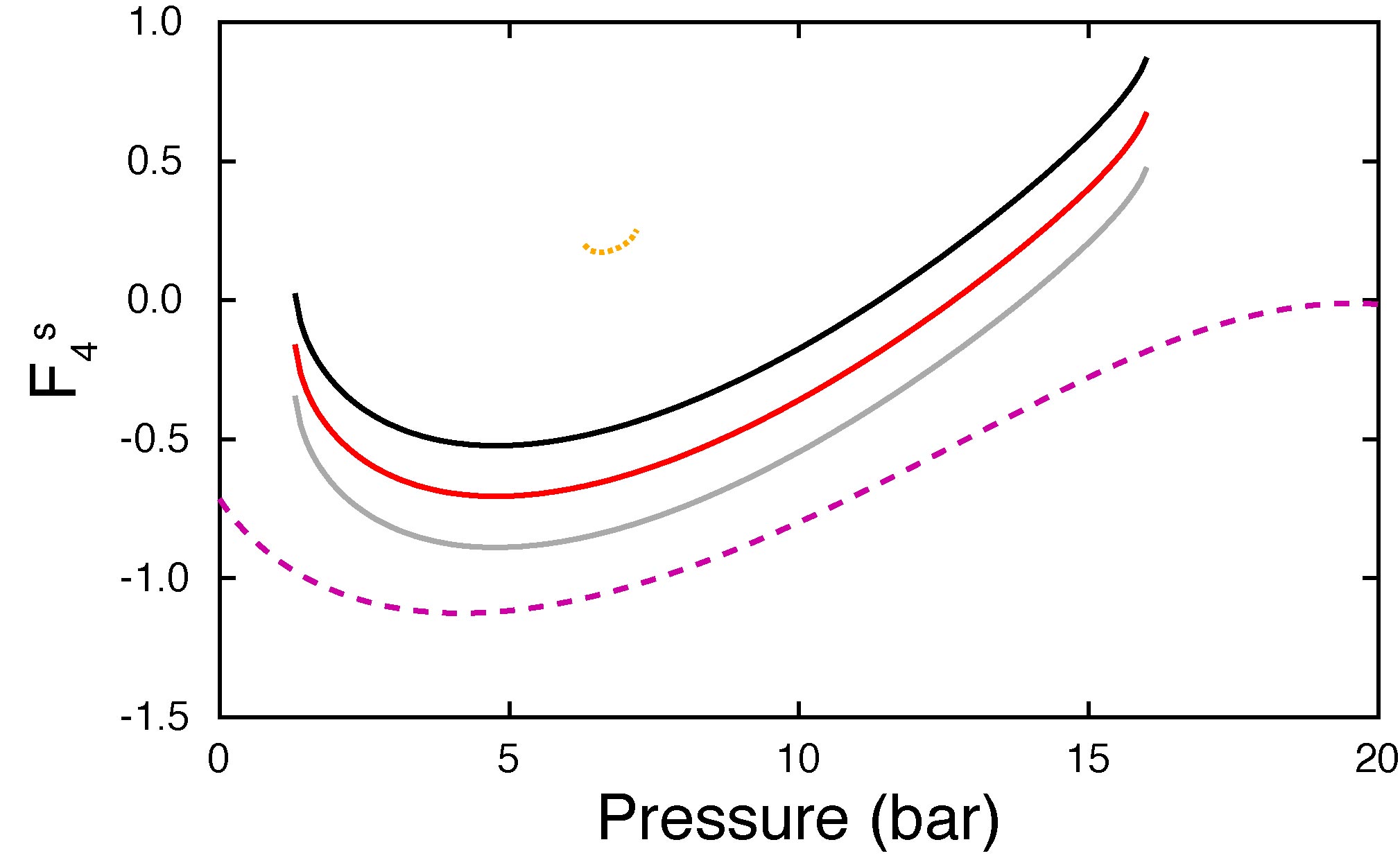}}
\caption{\label{F4s}(Color on-line).  $F_4^s$ calculated from Eq.~\ref{predJ4}
using $\Omega_{2\Delta}$-mode energies and $x_3^{-1}$ from RSQ-mode energies
\cite{Fra89,Fra90} (red curve).   $F_2^a$ is not particularly well known
\cite{Hal90} and alteration of $F_2^a$ by $\pm 0.2$ yields $x_3^{-1}$, which
give the black and grey curves, respectively.  The purple dashed curve results
from increasing $T_c$ by 1\%, and the orange dotted curve results from
decreasing $T_c$ by 1\%.}
\end{figure}

These values for $x_3^{-1}$ in Fig.~\ref{x3inv} are inconsistent with the
values we extracted from measurements of the frequency and Zeeman splitting of
the ISQ-mode \cite{Dav06,Dav08}.   However, as was noted previously, the effect
of interactions on the ISQ mode frequency, and possibly the Zeeman splitting of
this mode,  are likely affected by strong coupling corrections which are of
comparable magnitude \cite{Dav06,Dav08}.   On the other hand, the effect of
$x_3^{-1}$ on the RSQ-mode energies as reported by Fraenkel \emph{et al.}
\cite{Fra89,Fra90} is an order of magnitude
    larger than for the ISQ-mode energies and as a consequence, they are
possibly less affected by strong-coupling corrections.  Therefore, using the
RSQ-mode values of $x_3^{-1}$ and using the $\Omega_{2\Delta}$-mode energies,
it is possible to give possible values for $F_4^s$ from Eq.~\ref{predJ4}.
These values are given as the red curve in Fig.~\ref{F4s}.  Unfortunately, the
relevant Fermi liquid parameter, $F_2^a$, for calculation of
$x_3^{-1}$ from the RSQ-mode energies is not  well known \cite{Hal90}.
Increasing and decreasing
$F_2^a$ by 0.2 yields $x_3^{-1}$ given by the black and grey curves
respectively in Fig.~\ref{F4s}.  The purple dashed curve results from
increasing
$T_c$ by 1\%, and the orange dotted curve results from decreasing $T_c$ by 1\%.
Landau's formulation of interactions in
$^3$He in terms of a strongly converging expansion in Legendre polynomials with
coefficients, $F_l$,  would lead us to expect that the Fermi liquid parameters,
$F_l$, should decrease rapidly as $l$ increases.  According to the
    analysis above the absolute values of  $F_4^s$ might be comparable to, or
larger than, those of $F_2^s$ \cite{Hal90}.

\section{Conclusions}

	We have shown that low temperature pressure sweeps are a high 
resolution
technique to study the energy dependence of longitudinal and  transverse sound
in superfluid $^3$He-B, without the complicating effects of thermal damping
that occur in temperature sweeps.  This allows measurement of the absolute
values for the phase velocity of both longitudinal and transverse sound as a
function of energy.  These measurements reveal the coupling of transverse
sound to a collective mode near the pair-breaking edge, the $2\Delta$-mode,
which is likely the $J=4^-$ mode.  We use the
$2\Delta$-mode energies, in combination with the attenuation of longitudinal
sound, to set lower limits on the energy of pair-breaking.  We conclude that
pair-breaking never falls below the value given by weak-coupling BCS theory.
Additionally, we use the energies of this mode to estimate the strength of
sub-dominant $f$-wave pairing interactions.

	Future experiments with magnetic fields should bring the energy of
pair-breaking below that of the $2\Delta$-mode, because of the large
difference in their Zeeman splittings \cite{Dav08b,Sch81}.  This would allow
direct measurement of the pair-breaking energy using transverse sound, which
could then be extrapolated to zero magnetic field.  Direct measurement of the
energy of pair-breaking, with the precision of high frequency transverse
acoustics, can provide narrow constraints on the \emph{absolute} temperature
scale.

\section*{Acknowledgments}

	We acknowledge support from the National Science Foundation, 
DMR-0703656 and
thank J.A. Sauls, C.A. Collett, W.J. Gannon and S. Sasaki for useful
discussions.

\end{document}